\documentclass[aps,prl,twocolumn,endfloats]{revtex4b4}
\usepackage{bm}

\begin{document}

\title{Diffraction-free and dispersion-free pulsed beam propagation in dispersive media}
\author{Miguel A. Porras}
\affiliation{Departamento de F\'\i{}sica Aplicada. Escuela T\'ecnica Superior de
Ingenieros de Minas. Universidad Polit\'ecnica de Madrid. Rios Rosas 21. E-28003
Madrid. Spain}

\begin{abstract}
Pulsed Bessel beams of light propagating in free-space experience diffraction effects
that resemble those of anomalous dispersion on pulse propagation. It is then shown
that a pulsed Bessel beam in a normally dispersive material can remain diffraction-
and dispersion-free due to mutual cancellation of diffraction and group velocity
dispersion. The size of the Bessel transversal profile for localized transmission is
determined by the dispersive properties of the material at the pulse carrier
frequency.
\end{abstract}

\maketitle

Two of the biggest obstacles to the transmission of localized electromagnetic energy
over large distances are diffraction and material dispersion. Generally speaking,
diffraction makes waves to spread transversally to the intended propagation direction,
and dispersion temporally (longitudinally). Many methods have been proposed and
experientally demonstrated, to diminish, even eliminate either diffraction spreading
in free-space, by using diffraction-free Bessel beams, \cite{DU87} and their
generalizations, focus wave modes of various types, \cite{ZI89} optical missiles,
\cite{PO00} or dispersion spreading effects in dispersive media, by exploiting the
nonlinear properties of the medium, \cite{AG95} the dispersive properties of
diffraction gratings, \cite{SZ96} suitably designed Bessel-X waves, \cite{SO96} or the
pseudo-dispersion-free behavior of specific pulse temporal forms. \cite{RO95}

In the propagation of a transversally and temporally localized wave in a dispersive
material, both diffraction and dispersion effects act together, and lead, in general,
to an enhanced deterioration of the wave depth of field. However, as shown in this
paper, it is also possible to play off diffraction against dispersion during
propagation of a pulsed beam: by suitably designing its transversal profile, the
produced diffraction effects cancel, to a great extent, dispersion spreading, and vice
versa, leading to dispersion-free and diffraction-free  localized propagation in the
dispersive medium. Specifically, diffraction changes in a pulse with Bessel
transversal profile \cite{LI98} (do not confuse with the more known nondiffracting
X-Bessel waves) \cite{LU92},  and temporal spreading due to normal material dispersion
mutually cancel if the transversal size of the Bessel profile is properly chosen.

It is possible to arrive at this result by thinking of diffraction of pulses as a
dispersive phenomenon. Whenever the pulse has a transversal profile, diffraction
causes its redder frequencies to spread at larger angles than its bluer frequencies,
and hence to propagate at different effective velocities along the beam axis. A
detailed investigation on the dispersive nature of free-space diffraction of pulses,
including the description of diffraction forerunners, can be found in Ref.
[\onlinecite{BE94}]. Here we consider the light disturbance $E(\bm x_\perp, t)=g(\bm
x_\perp)A(t)\exp(-i\omega_0 t)$, with $\bm x_\perp \equiv(x,y)$, representing a pulse
of carrier frequency $\omega_0$, envelope $A(t)$, and transversal profile $g(\bm
x_\perp)$, at the entrance plane $z=0$ of a dispersive material of refraction index
$n(\omega)$, which fills the half-space $z>0$. The spatial-frequency spectrum of the
transversal profile $g(\bm x_\perp)$ is $\hat g(\bm k_\perp)$, with $\bm
k_\perp=(k_x,k_y)$, and the temporal-frequency spectrum of the pulse temporal form is
$\hat A(\omega-\omega_0)$. The propagated disturbance $E(\bm x_\perp, z, t)$ at any
plane $z>0$ inside the material can be seen as the result of superposing the
monochromatic plane waves $\hat g(\bm k_\perp)\hat A(\omega-\omega_0)
\exp\left[-i\omega t +i \bm k_\perp\cdot \bm x_\perp + ik_z(\omega)z\right]$ emitted
by the source plane, of different frequencies $\omega$, wavevectors $[\bm k_\perp,
k_z(\omega)]$, with
\begin{equation}\label{KZ}
k_z(\omega)=\sqrt{k^2(\omega)-|\bm k_\perp|^2} ,
\end{equation}
and $k(\omega)=(\omega/c)n(\omega)$, and amplitudes $\hat g(\bm k_\perp)\hat
A(\omega-\omega_0)$. These monochromatic plane waves are homogeneous if $|\bm
k_\perp|<k(\omega)$, and evanescent otherwise. To perform this superposition, we first
sum, for convenience, all monochromatic plane waves of different frequencies $\omega$
but same value of $\bm k_\perp$,
\begin{equation}\label{PUL}
E_{\bm k_\perp}(z,t)= \frac{\hat g(\bm k_\perp)}{2\pi}\int_{-\infty}^{\infty} d\omega
\hat A(\omega-\omega_0)\exp[-i\omega t+i k_z(\omega)z],
\end{equation}
and then superpose these partial fields,
\begin{equation}\label{SUM}
E(\bm x_\perp,z,t)=\frac{1}{(2\pi)^2}\int_{-\infty}^{\infty} d\bm k_\perp \exp(i\bm
k_\perp\cdot x_\perp) E_{\bm k_\perp}(z ,t).
\end{equation}
In this way, the propagated the pulsed beam appears as the superposition of many
pulses $E_{\bm k_\perp}$ associated to the different spatial-frequencies $\bm k_\perp$
of the initial transversal profile. The propagation of these subpulses is dispersive,
not only in dispersive materials, but also in free space [$k(\omega)=\omega/c$], since
their propagation constant $k_z(\omega)$ is a nonlinear function of $\omega$.

As in the usual theory of dispersive pulse propagation, we can expand $k_z(\omega)$
around the carrier frequency, $k_z(\omega) = k_{z,0} + k'_{z,0} (\omega-\omega_0) +
k^{\prime\prime}_{z,0}(\omega-\omega_0)^2/2 +\dots$ (where the prime sign denotes
differentiation with respect to $\omega$, and the subscript 0 evaluation at
$\omega_0$), to rewrite Eq. (\ref{PUL}), up to second order in dispersion, as
\begin{eqnarray}\label{ENV}
E_{\bm k_\perp}(z,t)&=& \hat g(\bm k_\perp)\exp(-i\omega_0 t + i k_{z,0}z) \nonumber
\\ &\times& \frac{1}{2\pi} \int_{-\infty}^\infty d\omega \hat A(\omega-\omega_0)
            \exp \left[\frac{i}{2} k_{z,0}^{\prime\prime}(\omega-\omega_0)^2z\right]
            \nonumber \\
   &\times& \exp\left[-i(\omega-\omega_0)\left(t-k'_{z,0}z\right)\right], \label{PRO2}
\end{eqnarray}
where, from Eq. (\ref{KZ}),
\begin{eqnarray}
k_{z,0}&=&\sqrt{k^2_0-|\bm k_\perp|^2}, \\ k'_{z,0}&=&\frac{k_0k'_0}{k_{z,0}}, \\
k_{z,0}^{\prime\prime} &=&\frac{k^3_0k^{\prime\prime}_0 -|\bm k_\perp|^2 [k^{\prime
2}_0+k_0k^{\prime\prime}_0]}{k_{z,0}^3}. \label{GVD}
\end{eqnarray}
When $|\bm k_\perp|<k_0$, $E_{\bm k_\perp}(z,t)$ is a propagating pulse, whose carrier
oscillations travel at the phase velocity $v_p=\omega_0/k_{z,0}$, while the envelope
does at the group velocity $v_g=(k'_{z,0})^{-1}$, at the same time that it broadens
due to the GVD of Eq. (\ref{GVD}).

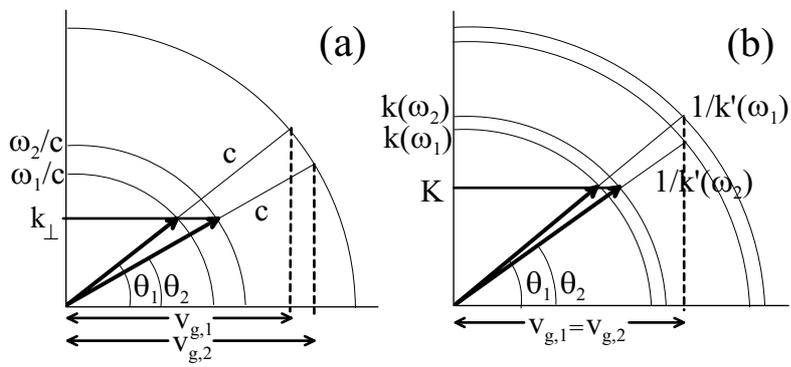
\begin{figure}
\caption{\label{UNO} (a) Illustration of the angular dispersion and the free-space
diffraction-induced anomalous dispersion in the group velocity
$v_g(\omega)=c\sqrt{1-(c|\bm k_\perp|/\omega)^2}<c$. Higher frequencies propagate at
greater group velocities. (b) Cancellation of material GVD dispersion with
diffraction-induced GVD. Provided that material dispersion is normal, i.e.,
$k'(\omega_1)<k'(\omega_2)$ for two close frequencies $\omega_1<\omega_2$, there
exists a particular value $K$ of $|\bm k_\perp|$ for which the effective group
velocities at $\omega_1$ and $\omega_2$ are equal.}
\end{figure}

In free-space, for instance, the phase and group velocities are $v_p=c/\sqrt{1-(c|\bm
k_\perp|/\omega_0)^2}>c$, $v_g=c\sqrt{1-(c|\bm k_\perp|/\omega_0)^2}<c$, \cite{BE94}
respectively, and the GVD $k^{\prime\prime}_{z,0} =-c|\bm
k_\perp|^2/\omega_0^3\sqrt{1-(c|\bm k_\perp|/\omega_0^2)^2}^3<0$ is anomalous.
Free-space dispersion originates from angular dispersion [see Fig. \ref{UNO}(a)]:
different frequencies $\omega$ composing the pulse $E_{\bm k_\perp}(z,t)$ propagate at
different angles $\sin\theta(\omega)=|\bm k_\perp|/(\omega/c)$ with respect to the $z$
axis, and then travel at different effective velocities along the $z$ direction. A
geometrical picture of $v_g$ and its dependence on frequency is shown in Fig.
\ref{UNO}(a). Free-space dispersion exists whenever there exists spatial-frequencies
$\bm k_\perp\neq 0$, i.e., the initial pulse has a transversal profile, and is
responsible for diffraction changes in the pulsed beam during propagation. Indeed, if
we neglect this kind of dispersion in Eq. (\ref{PUL}) by approaching
$\sqrt{(\omega/c)^2-|\bm k_\perp|^2}\sim\omega/c$,  Eq. (\ref{SUM}) would yield $E(\bm
x_\perp,z,t) = g(\bm x_\perp) A(t)\exp(-i\omega_0 t)$, i.e., the pulsed beam would
propagate without any change in free-space.

In a dispersive material, the total GVD of Eq. (\ref{GVD}) has two contributions,
originating from material dispersion and diffraction-induced dispersion. The
remarkable point here is that for normal material dispersion
($k^{\prime\prime}_{0}>0$), both types of GVD cancel mutually for pulses $E_{\bm
k_\perp}(z,t,)$ with
\begin{equation}\label{MAIN}
|\bm k_\perp|^2= K^2 \equiv\frac{k_0^3k^{\prime\prime}_0}{k^{\prime
2}_0+k_0k^{\prime\prime}_0},
\end{equation}
These pulses are not evanescent ($|\bm k_\perp|=K<k_0$), their propagating fields
being given, from Eq. (\ref{ENV}) with $k_{z,0}^{\prime\prime}=0$, by
\begin{equation}\label{ENV2}
E_{\bm k_\perp}(z,t)=\hat g(\bm k_\perp)\exp(-i\omega_0t+ik_{z,0}z)A(t-k'_{z,0}z),
\end{equation}
where $k_{z,0}=\sqrt{k_0^2-K^2}$ and $k'_{z,0}=k_0k'_0/\sqrt{k_0^2-K^2}$. Material and
diffraction-induced GVD cancelation is illustrated in Fig. \ref{UNO}(b).

Dispersion-free, diffraction-free pulsed beam propagation in a dispersive material can
then be achieved if the initial transversal profile contains only spatial frequencies
satisfying condition (\ref{MAIN}). The simplest example would be a single
spatial-frequency $\bm k_\perp$ of modulus $K$, but it does not represents a
transversally localized wave, but a tilted plane pulse. This is equivalent to the
result of Ref. [\onlinecite{SZ96}] for material GVD suppression by reflection of a
plane pulse in a grating of constant $\bm k_\perp$.  A second example, leading to
transversal localization, is the Bessel profile $g(\bm x_\perp) = J_0(K|\bm
x_\perp|)$, whose spectrum $\tilde g(\bm k_\perp) = \frac{2\pi}{|\bm
k_\perp|}\delta(|\bm k_\perp|-K)$ is an annulus of radius $K$. Indeed, Eqs.
(\ref{ENV2}) and (\ref{SUM}) for this spectrum yield the pulsed beam with nonspreading
envelope and transversal profile
\begin{equation}\label{PB}
E(\bm x_\perp,z,t)=J_0(K|\bm x_\perp|)A(t-k'_{z,0}z)\exp(-i\omega_0 t +i k_{z,0}z) ,
\end{equation}
as the propagated field of the initial disturbance $J_0(K|\bm x_\perp|) A(t)
\exp(-i\omega_0 t)$. We stress that this pulsed Bessel disturbance experiences
diffraction changes in free space, as shown in Ref. [\onlinecite{LI98}] (see also Fig.
2). The nondiffracting behavior of the pulsed Bessel beam in dispersive media can then
be explained by the mutual cancelation of diffraction and dispersion.

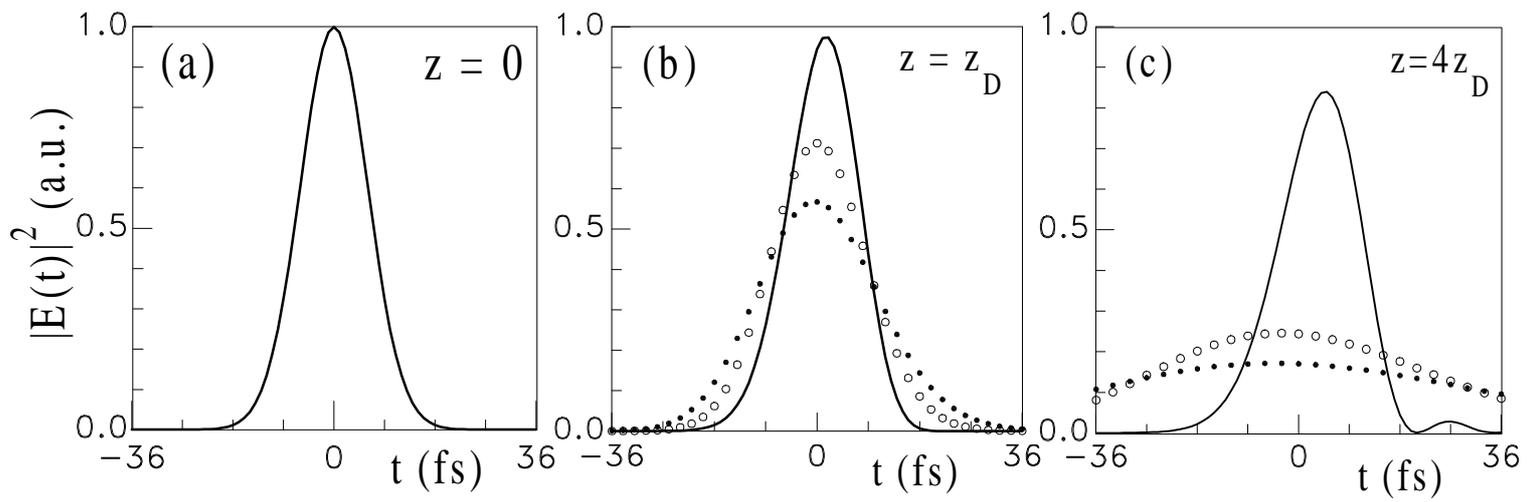
\begin{figure}
\caption{\label{DOS} Propagation of the pulsed Bessel disturbance $J_0(K|\bm
x_\perp|)\exp(-t^2/b^2)\exp(-i\omega_0 t)$ in fused silica (solid curves) and in
vacuum (dots), and of the pulsed plane wave $\exp(-t^2/b^2)\exp(-i\omega_0 t)$ in
fused silica (open dots). Numerical values of the parameters are $b=12$ fs, $K=
839.42$ mm$^{-1}$ and $\omega_0=1.9$ fs$^{-1}$. At this frequency the material
constants are $k_0=9193$ mm$^{-1}$, $k'_0=4881$ mm$^{-1}$fs and $k_0^{\prime\prime}=
21.78$ mm$^{-1}$fs$^2$. On-axis pulse forms at (a) $z=0$, (b)
$z=z_D=b_0^2/2|k_0^{\prime\prime}|= 3.3$ mm, and (c) $4z_D$. }
\end{figure}

To illustrate these results, Fig. \ref{DOS} shows the propagation of the pulsed Bessel
beam of Gaussian envelope $J_0(K|\bm x_\perp|)\exp(-t^2/b^2)\exp(-i\omega_0 t)$ in
fused silica (solid curves), with refraction index given by Selmeier relation. The
pulse duration and carrier frequency have been chosen arbitrarily to be $b= 12$ fs and
$\omega_0=1.9$ fs$^{-1}$ ($T_0 = 2\pi/\omega_0 = 3.3$ fs), respectively. Since
$k_0=9193$ mm$^{-1}$ , $k'_0=4881$ mm$^{-1}$ fs and $k_0^{\prime\prime}= 21.78$
mm$^{-1}$ fs$^2$ at this frequency, we have taken from  Eq. (\ref{MAIN}) $K=839.4$
mm$^{-1}$ for invariant propagation, yielding the beam width (first zero of the Bessel
profile) $2.404/K=2.864\;\mu$m, or about three times the carrier wavelength. For
comparison, we also show the propagation of the same pulse without transversal
modulation in fused silica (open dots), and of the same pulsed Bessel beam in
free-space (dots). It can be seen that the plane pulse in silica, under the only
action of dispersion, and the pulsed Bessel beam in free-space, under the effects of
diffraction only, have significantly spreaded at the dispersion length $z_D=
b_0^2/2|k_0^{\prime\prime}|= 3.3$ mm [Fig. \ref{DOS}(b)]. However, the pulsed Bessel
beam propagating in silica under the joint action of dispersion and diffraction does
not experience significant change up to $4z_D\simeq 13.2$ mm [Fig. \ref{DOS}(c)]. This
limitation is due to the total third-order dispersion $k^{\prime\prime\prime}_{z,0}$,
whose effect becomes noticeable at the third-order dispersion length
$b^3/2|k_{z,0}^{\prime\prime\prime}|=11.85$ mm.

Obviously, higher-order Bessel profiles, or the ``cos" beam (one dimensional version
of the Bessel beam) will also yield undeformable transmission. Spreading reduction is
also expected to occur with other transversal profiles having annular
spatial-frequency spectrum (though of finite thickness), as the Bessel-Gauss,
\cite{GO87} other windowed Bessel profiles, and the so-called elegant Laguerre-Gauss
beams \cite{PO01}. In these cases, invariant propagation will occur within the
diffraction-free distance \cite{GO87} (within which these profiles resemble the Bessel
one).

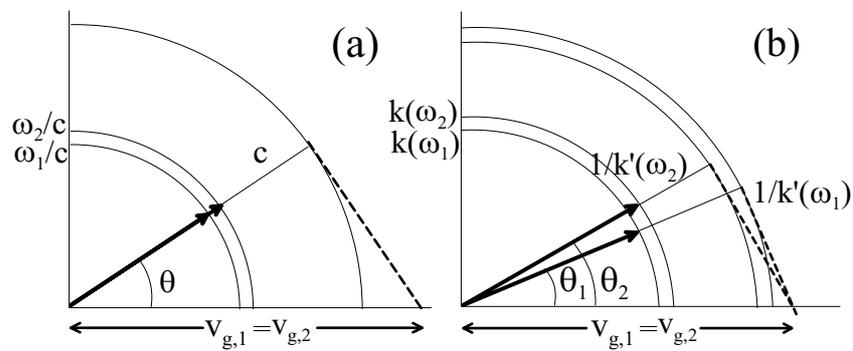
\begin{figure}
\caption{\label{TRES} (a) Illustration of the superposition scheme of a free-space
Bessel-X wave and its superluminal group velocity $v_g=c/\cos\theta >c$. All
frequencies travel at the same angle $\theta$, and hence $\bm k_\perp$ is proportional
to frequency. (b) Cancellation of material GVD by slightly distorting the cone angle
of the Bessel-X wave.}
\end{figure}

The above results must be clearly distinguished, despite some coincidences, from the
Bessel-X dispersionless propagation reported in Refs. [\onlinecite{SO96}], whose GVD
cancelation scheme is shown in Fig. \ref{TRES}, for comparison with Fig. \ref{UNO}.
Sonajalg's dispersionless pulse is built from a superluminal Bessel-X pulse [Fig.
\ref{TRES}(a)] having the nonseparable initial disturbance $\hat E(\bm x_\perp,
\omega)=\hat S(\omega)\delta[|\bm k_\perp|- k(\omega) \sin \theta]$ ($\bm k_\perp$
depends on frequency), instead from our separable pulsed Bessel beam [Fig.
\ref{UNO}(a)] $\hat E(\bm x_\perp, \omega)=\hat S(\omega)\delta[|\bm k_\perp|-K]$
($\bm k_\perp$ takes a fixed value). With the Bessel-X pulse, normal material GVD (for
instance) can be cancelled by slightly raising the cone angle $\theta$ of the
monochromatic Bessel beam components with increasing frequency [Fig. \ref{TRES}(b)].
Dispersion in the cone angle is supplied by an appropriate optical system, such as an
annular slit with frequency-dependent radius and a lens, an axicon, a lensacon plus a
telescope, depending on the dispersive material behind, or introducing some defocusing
in the lensacon. \cite{SO96} Here angular dispersion is inherent to the Bessel
profile: monochromatic Bessel beams of same size but different frequencies have the
different cone angles $\sin\theta(\omega)=K/k(\omega)$ [Fig. \ref{UNO}(b)]. It is also
to be noted that Sonajalg's pulse reduces to a pure nondistorted Bessel-X wave in the
limiting case of zero material GVD, whereas our pulsed Bessel beam degenerates into a
plane pulse ($K\rightarrow\infty$).

We have shown, in conclusion, that diffraction and material dispersion spreading
effects can cancel one to another during propagation of a pulsed beam in a dispersive
material, leading to dispersion-free, diffraction-free localized wave transmission, if
the transversal profile of the pulse is suitably selected and scaled. This result can
find application in ultrafast spectroscopy, large distance optical communications and
electromagnetic energy delivery systems.

\end{document}